\documentclass[allclo]{FBSart}
\usepackage{epsfig,psfig}
\newcommand{\cd}{\makebox[0.08cm]{$\cdot$}}
\newcommand{\sla}{\not\!}
\title{Renormalized non-perturbative scalar and fermion models in Covariant Light-Front Dynamics\footnote{Presented at Light-Cone 2004, Amsterdam, 16 - 20 August}}
\author{\underline{J.-F. Mathiot}*, V.A. Karmanov**, A. Smirnov**}
\institute{* Laboratoire de Physique Corpusculaire, Universit\'e Blaise Pascal,\\
 F-63177 Aubi\`ere Cedex, France\\
{**} Lebedev Physical Institute, Leninsky Prospekt 53, 119991 Moscow, Russia}
\runningauthor{J.-F. Mathiot}
\runningtitle{LC 2004}
\begin{document}
\maketitle
\begin{abstract}
Within the framework of the Covariant formulation of  Light-Front Dynamics, we develop a general
non-perturbative renormalization scheme, based on the Fock decomposition
of the state vector and its truncation. The explicit dependence of our
formalism on the orientation of the light front, defined by a light-like four
vector $\omega$, is essential in order to analyze the structure of the
counterterms needed to renormalize the theory. We illustrate our framework for scalar and fermion models.
\end{abstract}
\section{Introduction}
The knowledge of the hadron properties within the framework of QCD is one of the main issue in strong interaction physics.  Several approaches have been pursued in the last twenty
years, in particular lattice gauge calculations.  Among the
alternatives to these calculations, Light-Front Dynamics (LFD) is of
particular interest \cite{bpp}.  It has proven successful in many
phenomenological applications involving few-body systems in 
particle and nuclear physics. However, the application of LFD to field
theoretical calculations is still in its infancy. 
The main issue to
be solved is the renormalization procedure. In perturbative
calculations, the renormalization of the electron self-energy in QED,
in standard LFD, is already non-trivial in the sense that it involves
non-local counterterms.  This unpleasant feature is 
however a direct
consequence of the choice of a preferential direction, the $z$ axis,
in the determination of the quantization plane.  This can be well
understood in the Covariant formulation of Light-Front Dynamics
(CLFD) \cite{cdkm}, as shown in Ref. \cite{kms}.  
In this formulation, the state vector is defined on the light-front
plane given by the equation $\omega \cd x=0$, where $\omega$ is the 
four-vector with $\omega^2=0$. The particular case 
where $\omega=(1,0,0,-1)$ corresponds to standard LFD.

Our starting point is the general eigenstate equation for the state vector $\phi(p)$:
\begin{equation}\label{eq1}
\hat{P}^2\ \phi(p)=M^2\ \phi(p) \ .
\end{equation}
The momentum operator $\hat{P}_{\mu}$ is decomposed into two
parts: the usual free one, $\hat{P}^0_{\mu}$, and the interacting one
$\hat{P}^{int}_{\mu}$, given by:
\begin{equation}\label{kt3}
\hat{P}^{int}_{\mu}=\omega_{\mu}\int H^{int}(x)\delta(\omega\cd x)
\ d^4x=\omega_{\mu}\int_{-\infty}^{+\infty}
\tilde{H}^{int}(\omega\tau)\frac{d\tau}{2\pi} \ ,
\end{equation}
where $\tilde{H}^{int}$ is the Fourier transform of 
the interaction Hamiltonian. 
The state vector  $\phi(p)$ is then decomposed in Fock components. 
For convenience, it will be more appropriate to work with the vertex function defined by $\Gamma_i = (s-M^2) \phi_i$.
\section{Scalar model and non-perturbative renormalization}
For the simplified model we consider in this section, we take the
following interaction Hamiltonian:
\begin{equation}\label{eq4}
H^{int}=-g\psi^{2}(x)\varphi(x)\ ,
\end{equation}
where the scalar field $\psi$ with mass $m$ corresponds to  the
scalar "nucleon" and the
field $\varphi$ with mass $\mu$ corresponds to  the scalar "pion". In the approximation where we look for the dressing of the scalar "nucleon" by scalar "pions", only the mass of the nucleon is renormalized, and one should add a last term to the interaction Hamiltonian given by:
\begin{figure}[b]
\begin{center}
\psfig{figure=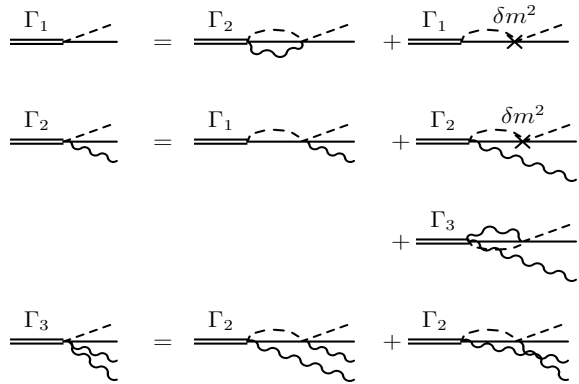,height=60mm,width=90mm}
\caption{Diagrammatic representation of the equation of motion for scalar particles.}
\end{center}
\end{figure}
\begin{equation}\label{eq5}
\delta H^{int}=-\delta m^2 \psi^{2}(x)\ .
\end{equation}
We have solved the eigenstate equation (\ref{eq1}) incorporating one, two and three-body components, as indicated on Figure 1. The mass counterterm is fixed from the condition that the eigenvalue is $M^2$. This is a non-perturbative condition. This counterterm is logarithmically divergent. Once this is done, the wave functions are finite and can be used to calculate all physical observables.
\section{Fermionic systems} 
\begin{figure}[bt]
\begin{center}
\psfig{figure=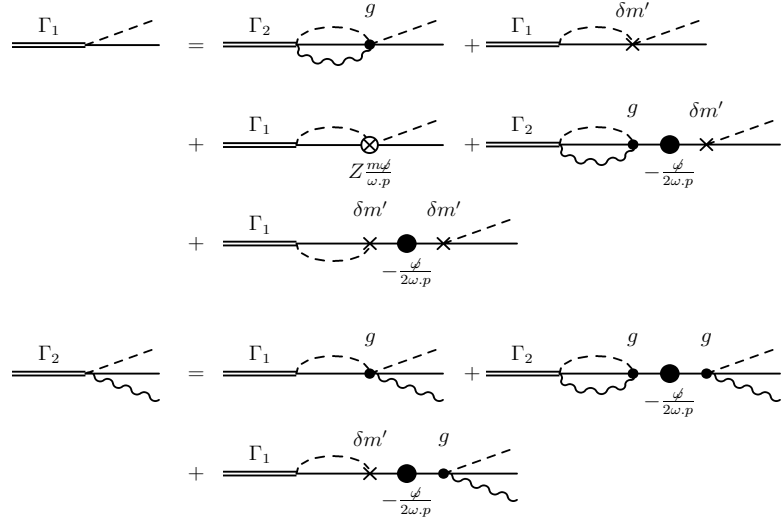,height=75mm,width=115mm}
\caption{Diagrammatic representation of the equation of motion for a fermion  coupled to a scalar boson. The counterterms $\delta m'$ and $Z$ are directly related to $\delta m$ and $Z_\omega$ \cite{kms}.}
\end{center}
\end{figure}
We now consider more realistic physical systems consisting of one fermion interacting with scalar or gauge bosons. In this study, we restrict ourself to one fermion and one fermion+one boson Fock components \cite{kms}. The analysis of the self energy in CLFD, together with the calculation of the two particle Green's function has led us to introduce a single new counterterm proportional to $\omega$
\begin{equation}
H_{\omega}(x)=-Z_\omega \bar{\psi}\frac{m \sla{\omega}}{i(\omega\cd\partial)}\ ,
\end{equation}
apart from the well known mass counterterm $\delta m$.
With this new contribution, the eigenstate equation (\ref{eq1}) can be rewritten according to Figure 2. Note the presence of contact interactions (denoted by black blobs) needed to get the right fermion propagator.

As already explained in \cite{cdkm}, the interacting part of the four-dimensional angular momentum operator is given by:
\begin{equation}\label{kt5}                                                                
\hat{J}^{int}_{\mu\nu}=\int H^{int}(x)(x_{\mu}\omega_{\nu} -x_{\nu}           
\omega_{\mu}) \delta(\omega\cd x)\ d^4x\ .                   
\end{equation}     
Its action on the state vector follows from the angular condition $\hat{J}^{int}_{\mu\nu}           \ \phi(p)= \hat{L}_{\mu\nu}(\omega)\phi(p)$, 
where $\hat{L}_{\mu\nu}$ is given by:
\begin{equation}\label{kt13}                                                    
\hat{L}_{\mu\nu}(\omega) =i\left(\omega_{\mu}                                   
\frac{\partial}{\partial\omega^{\nu}} -\omega_{\nu}                             
\frac{\partial}{\partial\omega^{\mu}}\right)\ ,                                 
\end{equation}                                                                  
This angular condition enables us to construct state vectors with well defined angular momentum.
For the case under consideration in this study, the general decomposition of the two-body component writes :                                                   %
\begin{equation}
\bar{u}(k_1)\Gamma_2 u(p)=b_1\bar{u}(k_1)u(p)+
b_2\frac{m}{\omega\cd p}\bar{u}(k_1)\sla{\omega}u(p)\ ,
\label{eq7}
\end{equation}
while the one body component $\Gamma_1$ is simply a constant. The three scalars $\Gamma_1,b_1,b_2$ are thus easily extracted from the eigenstate equation (\ref{eq1}) as explained in Ref.\cite{kms}. Since the residue of the two-body wave function should not depend on the arbitrary position of the light front plane at $s=M^2$, this condition fixes unambiguously, and non-perturbatively, the counterterm $Z_\omega$, while the mass counterterm $\delta m$ is fixed from (\ref{eq1}). This solves the problem. We have considered both a scalar boson or a gauge boson (either in the Light Cone gauge or in the Feynman gauge).
\section{Conclusion}
The explicitly covariant formulation of LFD is a very convenient tool in order to analyze the structure of the counterterms needed to renormalize bound state systems in a field theoretical description, and in particular their dependence on the orientation of  the light front plane. From the experience gained by these  first applications, we can propose a well defined strategy to address more realistic systems, like QED :
\begin{itemize}
\item The mass counterterm is fixed from the eigenvalue equation $P^2 = M^2$.
\item The $\omega$-dependent counterterm, $Z_\omega$, is fixed from the requirement that the two-body bound state wave function should be independent of the position of the light front at $s=M^2$.
\item All the counterterms depend on the number of Fock components which is considered. However, these counterterms can be fixed successively for the $N=1,2,...$ case from the same above physical requirements.
\end{itemize}
This strategy, although it never relies on perturbation theory, is completely consistent with it, as we explicitly shown when we restrict the Fock state to one fermion and one fermion+one boson, as it should. We shall  extent in the near future our study to calculate form factors, and extent our numerical calculation to include more Fock state components.
\end{document}